\documentclass[sigconf]{acmart}

\usepackage{booktabs} 
\usepackage{graphicx}
\usepackage{placeins}
\usepackage{balance}
\usepackage{xspace}
\usepackage{paralist}

\colorlet{color01}{red!10!white}
\colorlet{color02}{orange!10!white}
\colorlet{color03}{green!10!white}
\colorlet{color04}{blue!10!white}
\colorlet{color05}{magenta!20!white}
\colorlet{color06}{black!15!white}
\colorlet{color07}{yellow!10!white}
\colorlet{color08}{white}
\colorlet{color1}{red!80!black}
\colorlet{color1a}{red!50!white}
\colorlet{color2}{orange!80!black}
\colorlet{color2a}{orange!50!white}
\colorlet{color3}{green!50!black}
\colorlet{color3a}{green!90!black}
\colorlet{color4}{blue!80!black}
\colorlet{color4a}{blue!50!white}
\colorlet{color5}{magenta!80!black}
\colorlet{color5a}{magenta!50!white}
\colorlet{color6}{black!80!black}
\colorlet{color6a}{black!50!white}
\colorlet{color7}{white!30!black}
\colorlet{color7a}{black!50!white}
\colorlet{color8}{yellow}
\colorlet{color8a}{black!50!yellow}

\copyrightyear{2025}
\acmYear{2025}
\setcopyright{cc}
\setcctype{by-nc}
\acmConference[MM '25]{Proceedings of the 33rd ACM International Conference on Multimedia}{October 27--31, 2025}{Dublin, Ireland}
\acmBooktitle{Proceedings of the 33rd ACM International Conference on Multimedia (MM '25), October 27--31, 2025, Dublin, Ireland}\acmDOI{10.1145/3746027.3758167}
\acmISBN{979-8-4007-2035-2/2025/10}

\usepackage{booktabs}       
\usepackage{amsfonts}       
\usepackage{nicefrac}       
\usepackage{microtype}      
\usepackage{lipsum}
\usepackage{graphicx}
\usepackage{paralist}
\usepackage{caption}

\author{Aleksandr Farseev}
\email{farseev@itmo.ru}
\affiliation{%
   \institution{ITMO University}
   \institution{SoMin.ai Research}
   \city{Singapore}
   \country{Singapore}}

\author{Marlo Ongpin}
\email{marlo@somin.ai}
\affiliation{%
   \institution{SoMin.ai Research}
   \city{Singapore}
   \country{Singapore}}

\author{Qi Yang}
\email{yang@somin.ai}
\affiliation{%
   \institution{SoMin.ai Research}
   \city{Singapore}
   \country{Singapore}
   }

\author{Ilia Gossoudarev}
\email{goss@itmo.ru}
\affiliation{%
   \institution{Faculty of Software Engineering and
Computer Systems, ITMO University}
   \city{Saint Petersburg}
   \country{Russia}
   }

\author{Yu-Yi Chu-Farseeva}
\email{joy@somin.ai}
\affiliation{%
   \institution{SoMin.ai Research}
   \city{Singapore}
   \country{Singapore}
   }

\author{Sergey Nikolenko}
\email{sergey@logic.pdmi.ras.ru}
\affiliation{%
   \institution{ISP RAS Research Center for Trusted Artificial Intelligence}
   \city{Moscow}
   \country{Russia}
   }

\def\somon{MindFuse\xspace}

\title{MindFuse: Towards GenAI Explainability \\ in Marketing Strategy Co-Creation}

\begin{document}

\begin{abstract}
The future of digital marketing lies in the convergence of human creativity and generative AI, where insight, strategy, and storytelling are co-authored by intelligent systems~\cite{farseev2025aiagencies}. We present \somon, a brave new explainable generative AI framework designed to act as a strategic partner in the marketing process. Unlike conventional LLM applications that stop at content generation, \somon fuses CTR-based content AI-guided co-creation with large language models to extract, interpret, and iterate on communication narratives grounded in real advertising data.
\somon operates across the full marketing lifecycle: from distilling content pillars and customer personas from competitor campaigns to recommending in-flight optimizations based on live performance telemetry. It uses attention-based explainability to diagnose ad effectiveness and guide content iteration, while aligning messaging with strategic goals through dynamic narrative construction and storytelling. 
We introduce a new paradigm in GenAI for marketing, where LLMs not only generate content but reason through it, adapt campaigns in real time, and learn from audience engagement patterns. Our results, validated in agency deployments, demonstrate up to 12$\times$ efficiency gains, setting the stage for future integration with empirical audience data (e.g., GWI, Nielsen) and full-funnel attribution modeling. \somon redefines AI not just as a tool, but as a collaborative agent in the creative and strategic fabric of modern marketing. In the end of the paper, we also provide a forward-looking forecast on how platforms like MindFuse and Google DeepMind's ACAI are likely to shape the future of marketing agencies. These systems will not only reset client expectations toward greater transparency, speed, and personalization, but will also redefine the skillsets demanded from the new generation of marketers. As hybrid agencies emerge—blending creative storytelling with data science and AI engineering—the competitive landscape will increasingly hinge on talent. In this new environment, professionals will be expected to pair human imagination with technical fluency, while agencies will need to reinvent themselves as AI facilitators and curators of brand authenticity amidst the ongoing talent war led by platforms such as Meta.
\end{abstract}

\maketitle

\section{Introduction}
The online marketing ecosystem is undergoing a transformation driven by an exponential increase in the volume and complexity of multimedia content. From advertising libraries and social media to consumer behavior data, marketers today navigate a turbulent digital environment that demands not only speed and precision but also strategic depth and creativity. 
The primary challenge lies in synthesizing insights at scale without sacrificing interpretability \cite{_www25, _yang2023against, _yang2025fusing, _yang2024somonitor} or intent, which has been considered to be a major bottleneck of traditional machine learning in the marketing \cite{_farseev2018somin,_farseev2021somin, _huang2023socraft, _farseev2023under, _farseev2024somonitor} and, more specifically, AI-driven user profiling  \cite{_farseev2015harvesting,_yang2020know,_yang2023just, _yang2021two, _samborskii2019whole, _farseev2015cross, _farseev2014cross, _farseev2016bbridge, _farseev2017tweetfit, _buraya2017towards, _nie2017learning, _farseev2017cross, _farseev2017tweet, _farseev2017360, _buraya2018multi, _farseev2020understanding, _yang2022we, _yang2022personality, _chowdhury2017automatic, _samborskii2018person} domains.

Recent breakthroughs in LLMs have offered promising pathways for tackling these challenges. These models have demonstrated success across a range of domains, from low-resource translation and program synthesis to engineering and control systems~\cite{enis2024llm,chen2021evaluating,jiang2023mistral,kevian2024capabilities}. Their architecture continues to evolve, with newer Transformer variants such as Longformer~\cite{beltagy2020longformer}, Linformer~\cite{wang2020linformer}, and Mamba-style state-space models~\cite{gu2024mamba,qu2024surveymamba} drastically extending context lengths and reducing inference complexity~\cite{DBLP:journals/corr/VaswaniSPUJGKP17,ma2023mega}. These advances enable the direct analysis of large-scale text and multimedia datasets—a crucial requirement in marketing.

Furthermore, LLMs now possess multimodal capabilities~\cite{tang2024video,gpt4o}, allowing them to jointly process image, video, and audio alongside text. These systems can also be integrated with generative diffusion models~\cite{NEURIPS2020_4c5bcfec,Rombach_2022_CVPR} to generate campaign-ready media, making them increasingly relevant for advertising content production~\cite{hong2022cogvideo,liu2024sora}.

Despite these capabilities, a fundamental gap remains in applying LLMs to strategic marketing workflows. While they excel at generating content, LLMs perform inconsistently in \textit{content understanding}—a critical task in identifying high-impact ads and strategising future campaigns~\cite{zhang2023siren,chen2023hallucination}. This shortcoming is especially problematic given that industry-grade content assessment models used by platforms such as Meta~\cite{fb-ctr-practical-lessons}, Google~\cite{goog-ctr-view-from-trenches}, and Alibaba~\cite{baba-rep-learning-ctr} are proprietary and inaccessible to most marketers, and although there was a growing body of research in ad understanding~\cite{savchenko-etal-2020-ad}, before the advent of multimodal LLMs it had been insufficiently powerful to actually be practical.

To address this, we propose MindFuse, a Generative AI framework that reimagines the role of intelligent systems in marketing—not merely as tools for automation, but as co-strategists. MindFuse integrates a CTR-based prediction model with generative LLMs to power a novel workflow for campaign ideation, audience segmentation, and team briefing. Crucially, it shifts from static analysis to dynamic persona mining and narrative generation, effectively constructing high-quality content briefs aligned with market needs.

This system builds on recent advances in explainable and actionable AI~\cite{_yang2023against,ouyang2022training,openai2023gpt4}, using clustering techniques and LLM to extract communication personas and thematic challenges from large-scale ad corpora. These are synthesized into narrative briefs—compact, strategy-ready stories that marketers can use for content calendars, creative direction, or campaign pitches.

MindFuse is a novel brave idea, which is grounded in prior work on recommender systems and user modeling, which traditionally relied on topic models~\cite{10.1145/2615569.2615680}, convolutional networks~\cite{savchenko-etal-2020-ad}, and handcrafted user profiles~\cite{_buraya2018multi}. Our approach moves beyond these paradigms, offering LLM-powered, explainable pipelines that allow human strategists to collaborate with AI on equal footing. By automating the transformation of ad performance data into creative strategy, MindFuse envisions a new generation of AI systems that supplant human intuition in marketing.

\section{Content pillar identification}\label{sec:pillars}
Digital marketers routinely face the challenge of processing vast and constantly updating volumes of creative assets—ranging from advertisements and campaign messages to competitor materials and user interactions. In such a high-frequency, ``always-on'' environment, manual analysis of each asset or campaign becomes infeasible, creating a pressing need for scalable, automated interpretation tools.

To address this, we introduce a semantic layer within the MindFuse framework that leverages LLMs to extract structured insights from marketing content. These insights—referred to as \emph{content pillars}—form the foundation for downstream strategic analysis and narrative generation. They represent interpretable attributes such as customer needs, product value propositions, emotional appeals, and stylistic tone.

Our pipeline, begins by processing ad corpora through an LLM fine-tuned with domain-specific prompts. The model identifies structured semantic units in each creative asset (e.g., what need it targets, which product is featured), and stores these in tabular form. These content pillars are then aggregated and analyzed to reveal latent patterns in a brand's communication strategy—such as the dominant audience segments, key messaging themes, and emotional appeals.

\begin{figure}[!t]\centering
\includegraphics[width=\linewidth]{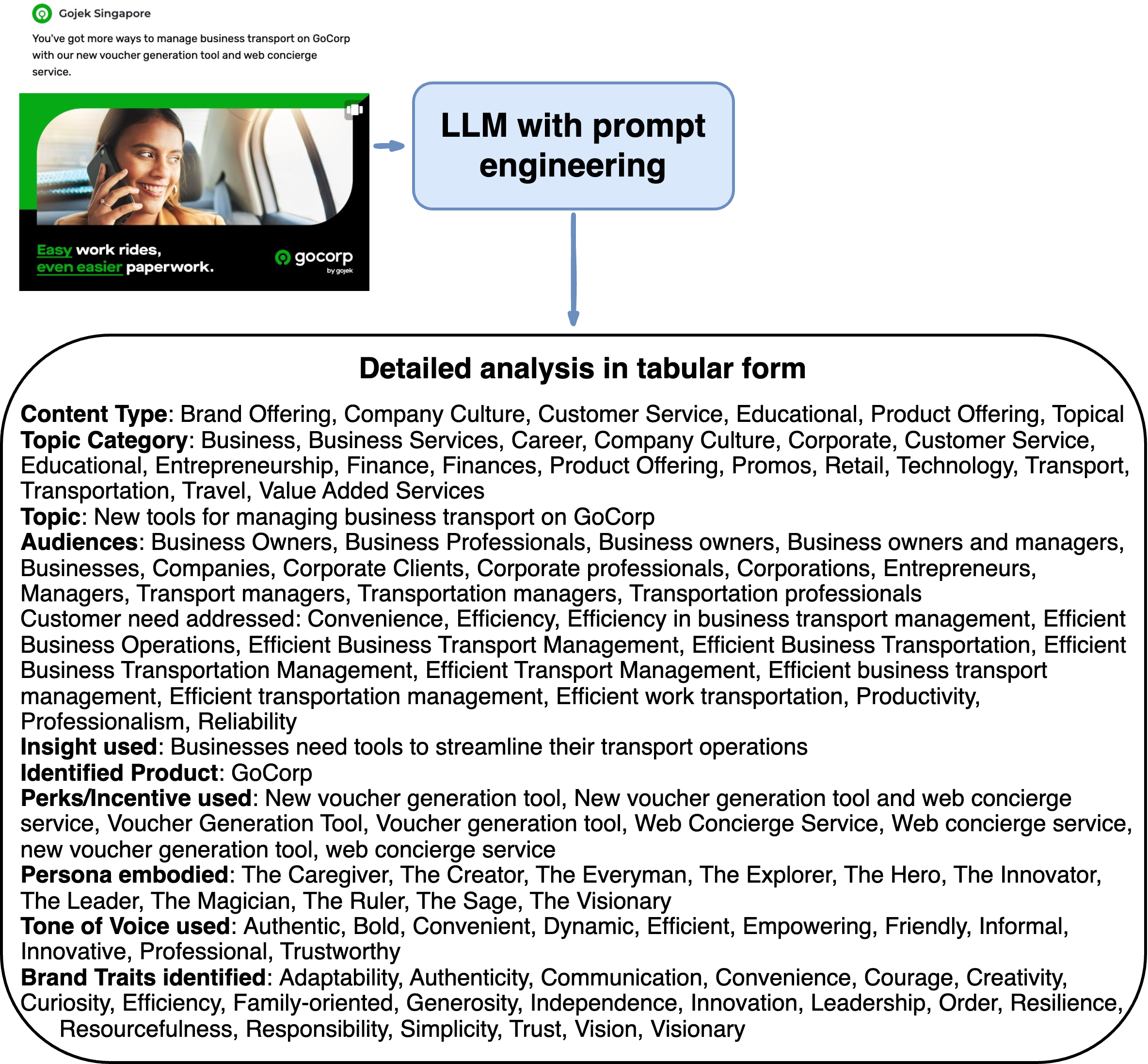}
\caption{Sample ad analysis via LLM-powered content pillar extraction.}
\label{fig:adanalysis}
\vspace{-1em}
\end{figure}

To demonstrate this process in action, we conducted an empirical analysis of advertising content from four major Singapore-based transportation companies. As shown in Figure~\ref{fig:adanalysis}, we collected batches of ads from the Facebook Ad Library, then parsed them through LLM prompts designed to simulate a strategic planner’s line of questioning—e.g., “What is the underlying human need targeted here?”, “What is the dominant product category?”, or “Which archetype is invoked in this message?”

The resulting outputs were standardized and highly interpretable, enabling downstream automation and large-scale comparison across campaigns. Unlike traditional manual content audits, which are time-intensive and limited in scope, our approach enables rapid, scalable analysis across thousands of assets—laying the groundwork for deeper persona and theme discovery, as described in Section~\ref{sec:perstheme}.

\section{Mining Communication Themes}\label{sec:perstheme}
\subsection{From Description to Strategic Inference}

Extracting surface-level descriptors such as customer needs and product categories—what we refer to as content pillars—is only the first step toward intelligent marketing systems. To truly augment strategy design, AI must engage in higher-order reasoning: discerning why certain messages resonate, who they target, and how they can be recombined to create new, context-aware narratives.

Yet most marketers operate without formal data science training, and current tools rarely support strategic ideation beyond dashboards and KPIs. This creates a gap between accessible insights and actionable planning. MindFuse bridges this gap through a unified pipeline that integrates content scoring and LLM-driven clustering to surface latent strategic elements—namely, customer personas and thematic challenges—from large ad corpora. These form the foundational elements of generative marketing storytelling, explored further in Section~\ref{sec:story}.

\subsection{Uncovering Personas from The Data}\label{sec:persona}

Customer personas are traditionally crafted through manual synthesis of demographics, survey data, and creative intuition. In MindFuse, we replace this manual bottleneck with a semi-automated process that uses unsupervised clustering and generative summarization. Using X-Means clustering~\cite{pelleg2000x} on ADA embeddings~\cite{neelakantan2022text} derived from the ``Audience'' content pillar, we identify meaningful groupings that reflect distinct psychological and behavioral marketing profiles.

\begin{figure}[!t]
\includegraphics[width=\linewidth]{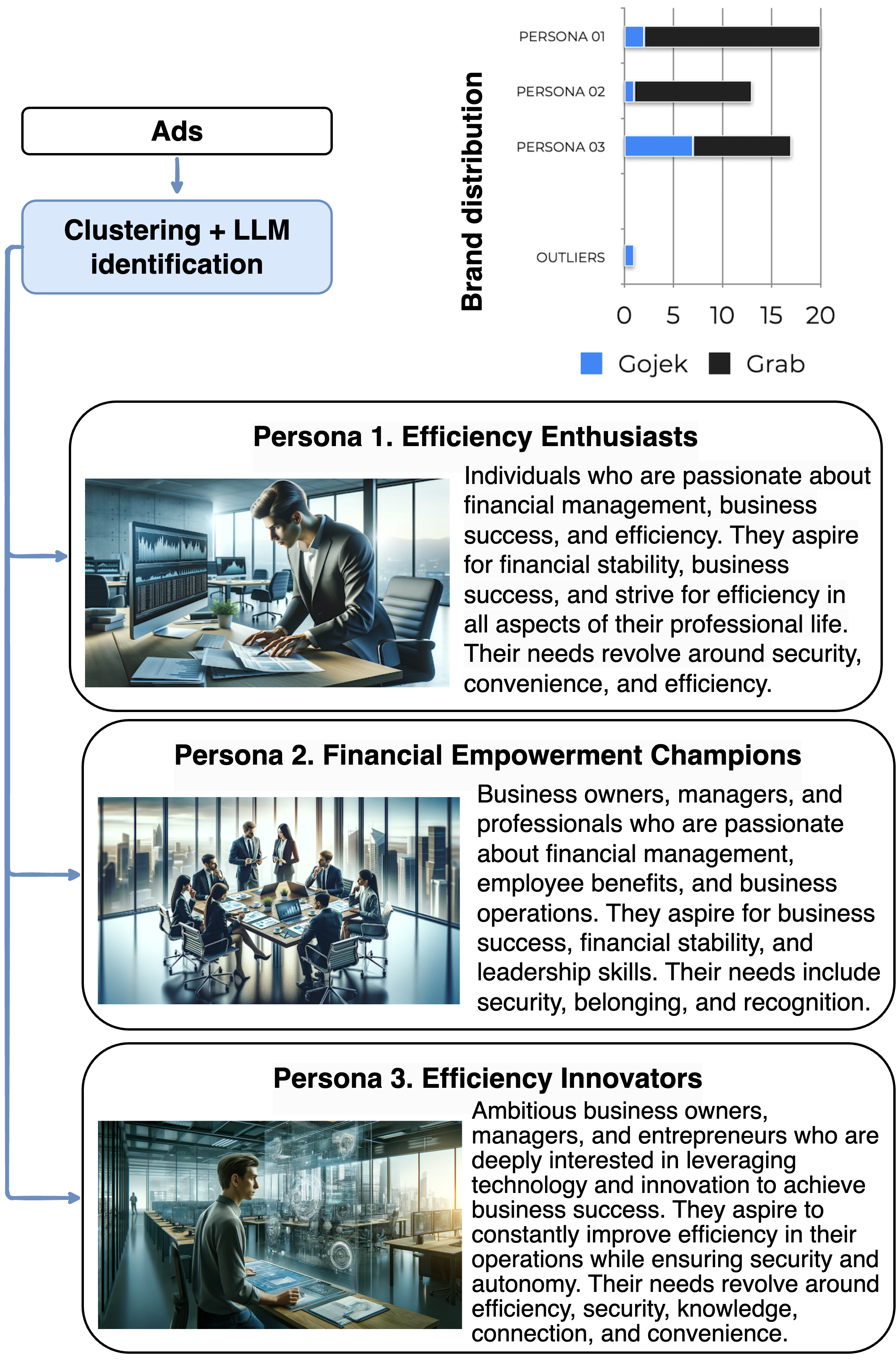}
\caption{MindFuse-generated customer personas based on semantic clustering of B2B ad content.}
\label{fig:personas}
\end{figure}

To ground the analysis in a competitive context, we collected 1,120 business-focused ads from Gojek and Grab, Singapore's two leading ride-hailing platforms, covering 2023–2024. These ads—part of a larger corpus of 5,967 content items—were pre-filtered to focus on B2B communication. Fig.~\ref{fig:personas} presents the three personas that emerged:
\begin{inparaenum}[(1)]
\item \emph{Efficiency Enthusiasts} (206 items): traditional-minded decision-makers focused on operational cost savings,
\item \emph{Financial Empowerment Champions} (144 items): advocates of equitable and flexible employee benefits, and
\item \emph{Efficiency Innovators} (707 items): technology-forward adopters seeking scalable digital solutions.
\end{inparaenum}

The resulting clusters are not just static categories. Through LLM-guided summarization and image generation, each persona becomes a narrative archetype that can be woven into campaign strategy. This synthesis transforms latent engagement patterns into design-ready story templates—machine-mined but human-usable.

\subsection{Mining Thematic Challenges}\label{sec:theme}

A persona alone is not enough to drive effective storytelling. Just as protagonists are defined by the challenges they face, marketing narratives require a complementary communication theme—often rooted in the implicit tension the campaign seeks to resolve.

To identify such themes, we applied a second round of X-Means clustering to the ``Insights'' content pillar, again using ADA embeddings~\cite{neelakantan2022text}. These clusters reflect recurring pain points, value propositions, or aspirations across campaigns. The LLM was prompted to label and describe each theme based on the dominant content features within the cluster.

\begin{figure}[!t]\centering
\includegraphics[width=.9\linewidth]{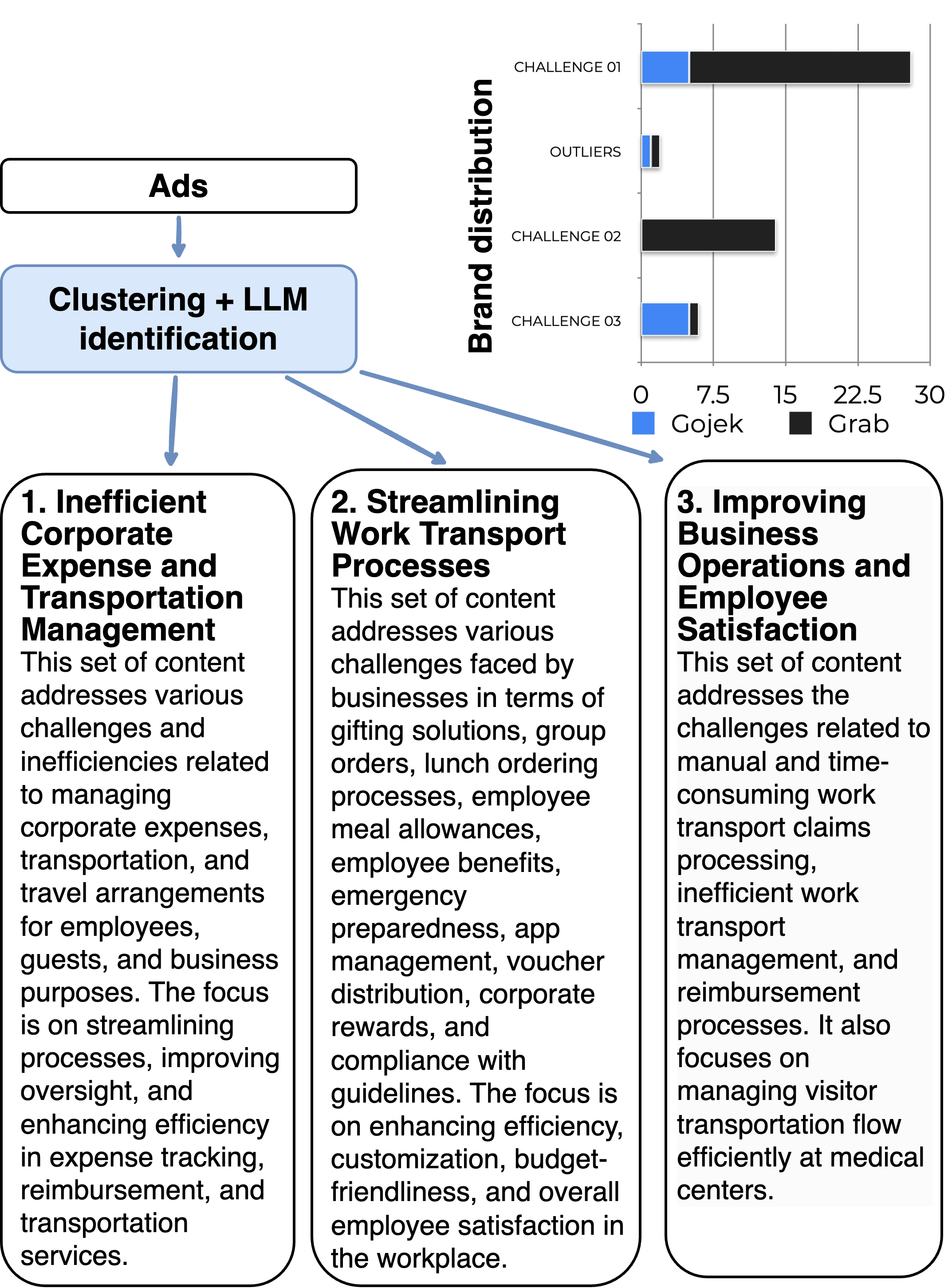}
\caption{Representative communication challenges extracted from insight-driven clustering.}
\label{fig:challenges}
\end{figure}

As shown in Fig.~\ref{fig:challenges}, the system surfaced three distinct challenge themes:
\begin{inparaenum}[(1)]
\item \emph{Inefficient Corporate Expense and Transportation Management} (672 items),
\item \emph{Improving Business Operations and Employee Satisfaction} (94 items), and 
\item \emph{Streamlining Work Transport Processes} (336 items).
\end{inparaenum}
These themes serve as strategic lenses through which brands frame their messaging. When paired with personas, they form complete narrative skeletons—fitting the archetypal “hero + conflict” structure—ready for story generation. Unlike static segmentations used in traditional marketing, these LLM-assisted groupings are fluid, data-driven, and continuously refreshable as campaign data evolves.

In summary, this dual-layer personas and themes mining demonstrates how MindFuse turns raw engagement data into generative storytelling assets. It moves beyond explainability toward a future where AI co-authors the narratives that define brand identity.

\section{Real-World Application}\label{sec:story}
\begin{figure}[!t]
\includegraphics[width=\linewidth]{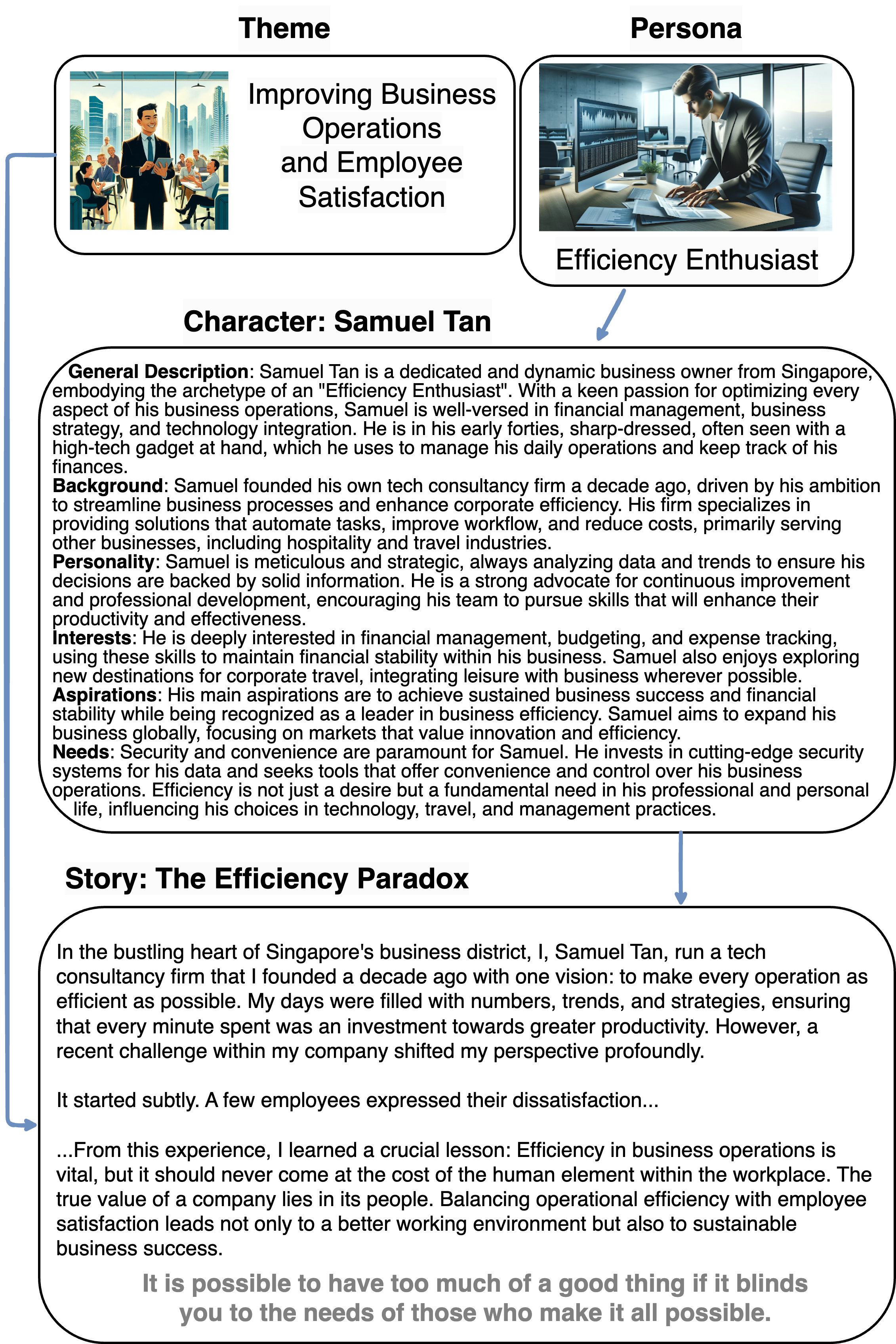}
\caption{LLM-generated campaign brief: combining persona + challenge into a strategic narrative.}
\label{fig:story-sample}
\vspace{-1em}
\end{figure}

MindFuse advances beyond analysis by generating strategic narratives that marketers can directly apply. Building on previously extracted personas and challenges, the system synthesizes these into personalized campaign stories using LLM prompts (Fig.~\ref{fig:story-sample}). This generative pipeline simulates real-world strategy design workflows—replacing manual brainstorming with data-driven creativity.

Based on detected campaign gaps, MindFuse can suggest targeting underused personas (e.g., “Efficiency Enthusiasts”) with overlooked challenges (e.g., “Streamlining Work Transport”). The LLM then generates a narrative—such as one about a fictional business owner, “Samuel Tan”—highlighting how the brand addresses both practical and emotional needs. These stories serve as high-quality content briefs, on par with those crafted by human strategists, yet produced in seconds. With a simple prompt, the story is distilled into a campaign insight, which, when paired with actual offerings like GoJek’s Corporate Car Hailing and framed by the selected challenge, forms the foundation of a compelling campaign idea—typically a task requiring days of agency work.

To evaluate the applicability of this approach in the real-world settings, we deployed MindFuse across four organizations—two creative agencies, one branding firm, and one consumer brand—evaluating time savings in core strategic tasks. As shown in Table~\ref{tab:time_savings_icde}, marketing teams reported 2×–5× reductions in task durations for content strategy, campaign planning, and audience research.

\begin{table}[!t]
\centering
\caption{Time savings from LLM-based narrative generation.}
\label{tab:time_savings_icde}
\begin{tabular}{lcc}
\toprule
\textbf{Task} & \textbf{Before (hrs)} & \textbf{After (hrs)} \\
\midrule
\multicolumn{3}{l}{\textbf{NEO360 (Performance Agency~\cite{neo3602025})}} \\
Social Media Proposals & 3.0 & 0.75 \\
Competitor Monitoring & 3.0 & 0.25 \\
Audience Research & 5.0 & 0.5 \\
Ad Brainstorming & 2.5 & 0.25 \\
\midrule
\multicolumn{3}{l}{\textbf{Blak Labs (Creative Agency)~\cite{blaklabs2025}}} \\
Campaign Planning & 5.0 & 2.0 \\
Content Planning & 10.0 & 3.0 \\
Brand Analysis & 10.0 & 1.0 \\
\midrule
\multicolumn{3}{l}{\textbf{Bit by Bit (Branding Agency)~\cite{bitbybit2025}}} \\
Content Strategy & 5.0 & 1.0 \\
\midrule
\multicolumn{3}{l}{\textbf{Mothercare (Consumer Brand)~\cite{mothercare2025}}} \\
Competitor Analysis & 5.0 & 2.0 \\
Ad Compilation & 5.0 & 1.0 \\
Customer Segmentation & 5.0 & 2.0 \\
\bottomrule
\end{tabular}
\end{table}

The observed efficiency gains—from 2.5× to 12×—demonstrate that LLM-generated campaign briefs are not only practical but transformative. MindFuse empowers marketers to move seamlessly from insight to execution, converting audience signals and ad performance data into actionable strategic narratives. This marks a significant shift toward co-creative AI in brand storytelling, redefining how marketing strategies are conceived and delivered.

\section{Content Scoring and Explainable Content Optimization}

Once strategic direction is set, teams invest heavily in generating ad visuals, copy, and campaign assets. However, empirical studies reveal a sobering fact: only around 20\% of ad creatives succeed in delivering strong campaign performance~\cite{farseev2024transparency}. The rest? They consume budgets with little return.

This inefficiency is not just a production problem—it’s a strategic blind spot.

To confront this challenge, we propose a novel integration of predictive modeling and explainability into the creative pipeline. Within the MindFuse framework, we introduce a Content Scoring Module, informed by Yang et al. ~\cite{yang2025fusing}, that evaluates advertising assets before they go live. Built upon the SODA architecture for CTR prediction and recommendation, this module goes further by offering interpretable attention-based heatmaps—enabling marketers to understand why certain creatives are likely to perform better than others.

This represents a shift from post hoc measurement to pre-launch optimization, towards data-driven creativity where insight and imagination converge to reduce waste and unlock impact at scale.

\subsection{Predictive Scoring with Explainable Visual Guidance}
Our system generates heatmaps from attention layers within the scoring model, highlighting regions most influential in driving predicted user engagement. In this study, we examined whether marketers can leverage these visual cues to optimize creative assets and improve campaign performance.

\begin{figure*}[!t]
\centering
\includegraphics[width=\textwidth]{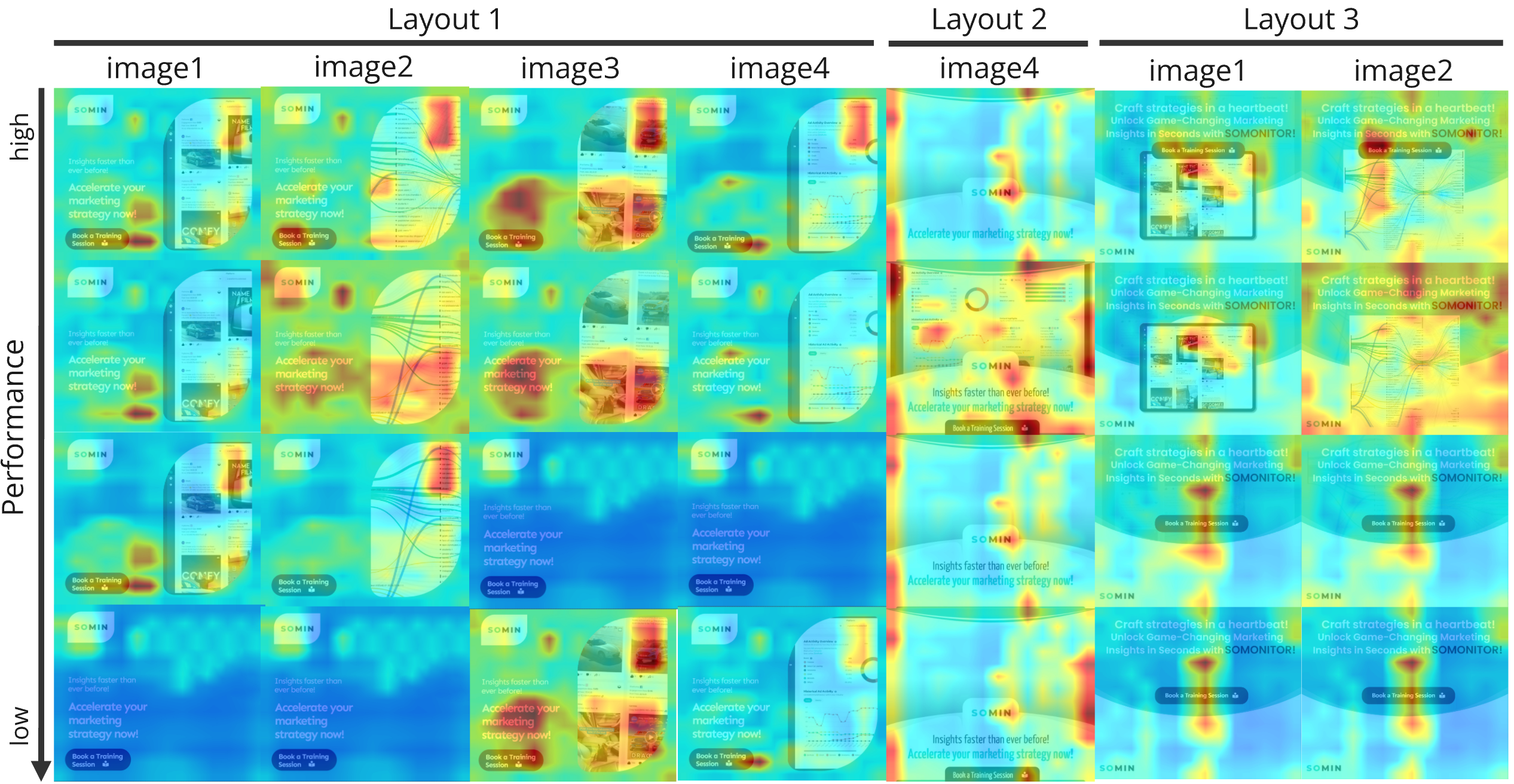}
\caption{Heatmaps visualizing high-attention regions in ad creatives.}
\label{fig:heatmap}
\end{figure*}

We deployed original and modified versions of ad creatives on the \emph{Meta} advertising platform to test the impact of removing heatmap-highlighted elements. Each creative underwent the following:
\begin{itemize}
\item ads are created using MindFuse-generated stories as described in the previous sections;
\item heatmaps are visualizations of the attention layers of the CTR prediction model~\cite{yang2025fusing} to highlight high-impact regions;
\item critical elements were removed iteratively to create variants;
\item all creatives were deployed under identical Meta campaign settings (Objective: Landing Page View Optimization; Budget, Targeting, Placements, etc. were equal for all experiments for an ``apple-to-apple'' comparison);
\item we recorded performance across 65,107 impressions, 2,243 clicks, and 1,162 LPVs.
\end{itemize}

\subsection{Experimental Findings}
\begin{table}[!t]
\centering
\caption{Performance degradation after object removal.}
\label{tab:evaluation}
\begin{tabular}{lcccc}
\toprule
\textbf{Layout} & \textbf{LPV} & \textbf{CTR-LPV} & \textbf{CTR} & \textbf{F1 Score} \\
\midrule
Layout 1 & 0.769 & 0.769 & 0.692 & 0.750 \\
Layout 2 & 0.751 & 0.751 & 0.250 & 0.500 \\
Layout 3 & 0.570 & 0.857 & 0.857 & 0.860 \\
\midrule
\textbf{Overall} & 0.708 & 0.792 & 0.667 & 0.740 \\
\bottomrule
\end{tabular}
\end{table}

Results in Table~\ref{tab:evaluation} confirm the utility of attention maps in identifying high-impact visual features:
\begin{itemize}
\item Removing the \textbf{primary visual element} reduced CTR by 58\%, showing its key role in user attention.
\item Eliminating the \textbf{CTA button} resulted in a 22\% drop in CTR, reflecting its engagement-driving power.
\item Replacing the \textbf{transparent background} led to a 64\% CTR drop---the most severe impact---indicating its importance for contrast and visual clarity.
\end{itemize}

\begin{figure*}[!t]
\centering
\includegraphics[width=\textwidth]{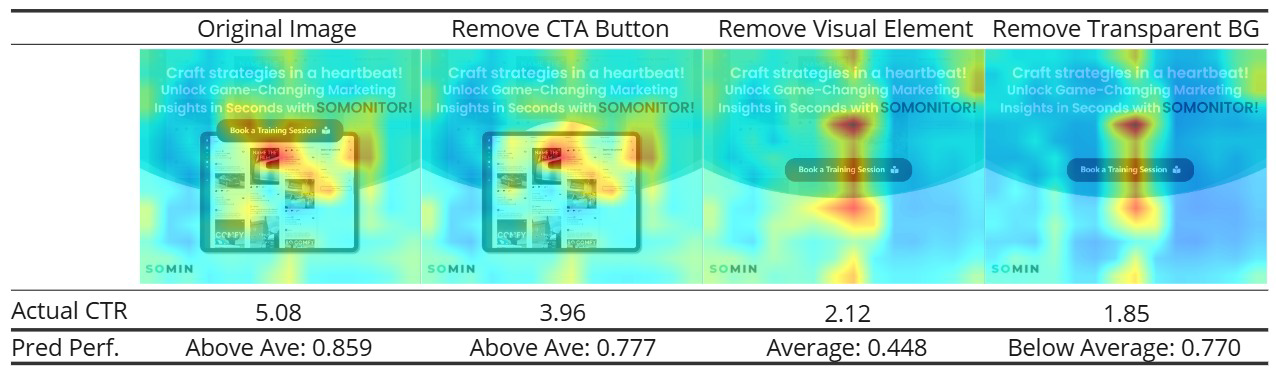}
\caption{Performance comparison across modified creatives.}
\label{fig:comparison_ads}
\end{figure*}

\subsection{Implications for Creative Strategy}
These results demonstrate that MindFuse's explainable content scoring approach, enhanced with explainability through heatmaps, not only forecasts performance but also provides practical guidance to optimize visual storytelling. Marketers can thus make data-driven design choices that preserve or enhance campaign impact.

This approach shifts content scoring from black-box prediction to an actionable co-creative process, enabling better-informed decision-making and more effective content iteration at scale.

\section{GenAI as a Performance Marketer}
\label{sec:optimization}

With high-performing creatives now selected through explainable and iterative predictive modeling, the next frontier lies in scaling campaigns under real-world constraints—namely, fixed creative pipelines, high media spend, and rigid approval cycles. In such environments, the traditional assumption that quality creatives alone drive results quickly breaks down. Even top-performing assets can stagnate if not aligned with evolving audience dynamics or strategic shifts.
We propose a paradigm where large language models function as autonomous performance strategists, augmenting campaign optimization in-flight. Rather than relying solely on manual intervention, these agents ingest live campaign telemetry, interpret micro-level content performance signals, and translate them into macro-level decisions on budget allocation, targeting refinement, and pacing strategies.
This approach reframes LLMs from content generators to adaptive, real-time interpreters of marketing performance—closing the feedback loop between creative impact and strategic execution in high-stakes campaign environments.

\subsection{Data and Performance Metrics}

Our system interprets campaign health across the following key metrics:
\begin{inparaenum}[(i)]
\item \emph{Reach}: Unique individuals exposed to the campaign.
\item \emph{Frequency}: Average ad exposures per user—critical to monitor fatigue.
\item \emph{Results}: Conversions, leads, or purchases.
\item \emph{Cost per Result (CPR)}: Conversion efficiency.
\item \emph{Spend}: Budget deployed.
\item \emph{CPM}: Cost per 1,000 impressions—market competitiveness proxy.
\item \emph{CTR}: Ad engagement rate.
\item \emph{CR: Click-to-View}: \% of users who land on the destination page.
\item \emph{CR: Click-to-Result}: \% of users completing the campaign goal.
\end{inparaenum}

\begin{figure*}[!t]
\includegraphics[width=\textwidth]{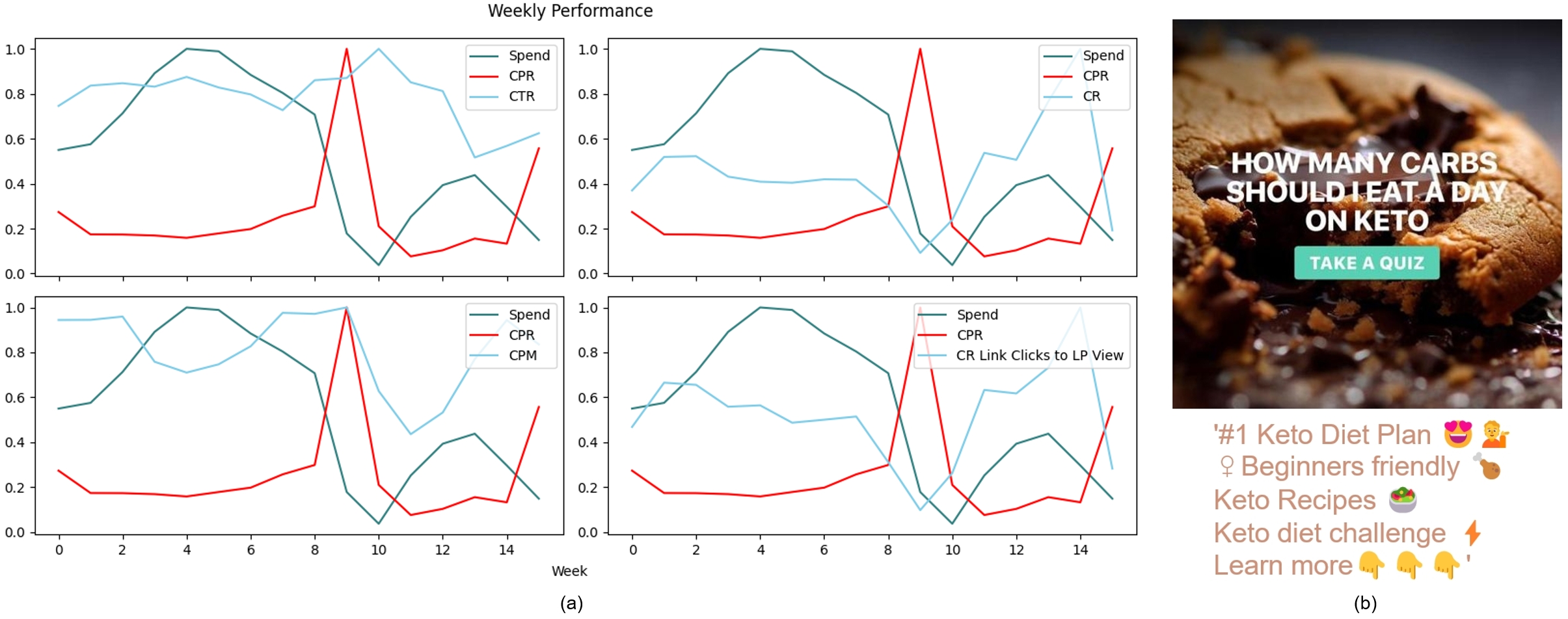}
\caption{Meta campaign data: (a) weekly trendline; (b) sample creative.}
\label{fig:trend}
\end{figure*}

We monitored a three-month Meta campaign (Oct 2023–Jan 2024), collecting metrics at three levels:
\begin{inparaenum}[(i)]
\item weekly—to smooth anomalies,
\item daily—for day-to-day variance, and
\item by creative—to evaluate asset-level impact.
\end{inparaenum}
To make the data LLM-readable, we resized the visuals to 512×512 and encoded them and transformed metrics into percentage changes to emphasize trends.

\subsection{Prompt Design for Campaign Performance Analysis}

LLM-driven optimization relied on a custom prompt structure: \textbf{Knowledge + Role + Task + Guiding Questions + Dat Input}.

\textbf{Knowledge:} Metric definitions (Reach, CPR, CTR, etc.) were embedded in the prompt for consistent interpretation.

\textbf{Role:} The LLM was cast as a senior performance marketing lead, expected to deliver decisive, insight-driven guidance—avoiding hedging language.

\textbf{Task:} The core task was CPR minimization. The LLM reviewed trends in CPR and Spend alongside CTR, CPM, and CR to identify strategic shifts.

\textbf{Guiding Questions:} The LLM answered:
\begin{enumerate}
\item How did CPR and Spend evolve?
\item What were the corresponding changes in CTR, CPM, and CR?
\item Were these changes causally connected?
\item Which secondary metrics influenced CPR the most?
\item What creative-level insights explain these shifts?
\item What actions should be taken?
\end{enumerate}

\textbf{Data Input (Tabular + Visual):} Prompts included tabulated performance data and ad creatives to enable both analytical and visual evaluation—critical for multimodal LLMs like GPT-4.

\subsection{Emergent Strategic Thinking in GenAI Campaign Analysis}
In the absence of ground-truth supervision or domain-specific fine-tuning, the LLM, prompted only with structured campaign outputs, exhibits signs of emergent strategic reasoning. Operating as a zero-shot performance analyst, the model autonomously navigates from signal detection to tactical recommendations—mirroring the thought process of a seasoned media strategist. The unfolding steps below illustrate the model’s capacity for interpretive abstraction and causal synthesis in the context of campaign analytics:

\textbf{Step 1: Salient Metric Disruption.}
The model independently identified volatility across key performance indicators—most notably, Cost Per Result (CPR) and Spend—without explicit cues, flagging instability in campaign dynamics.

\textbf{Step 2: Interpreting Efficiency and Pacing Variance.}
It quantified extreme disparities in CPR (ranging from \$27.68 to \$208.28) and total Spend (\$175.18 to \$4664.91), diagnosing inefficiencies in budget deployment and inconsistent cost-effectiveness.

\textbf{Step 3: Auxiliary Signal Correlation.}
Beyond primary metrics, the LLM tracked secondary signals such as CTR (peaking at 3.59\%) and a declining Conversion Rate—interpreting them as symptoms of audience fatigue. Concurrent CPM spikes (up to \$14.06) were read as evidence of heightened auction competition.

\textbf{Step 4: Proto-Causal Inference.}
Despite lacking causal training, the model connected declining CR and rising CPM as joint contributors to CPR escalation—demonstrating an emergent ability to reason through multi-variable interdependencies.

\textbf{Step 5: Narrative Synthesis.}
The LLM generated a temporal hypothesis of campaign behavior: initial targeting success, followed by mid-phase market saturation and late-stage creative fatigue. This phased interpretation was neither prompted nor pre-encoded—suggesting latent theorization capabilities.

\textbf{Step 6: Cross-Modal Creative Evaluation.}
Through textual analysis, the model affirmed audience-product alignment (e.g., Keto interest), but identified visual creative gaps—specifically a lack of compelling calls-to-action. It proposed A/B testing on visual assets, a task typically reserved for experienced human planners.

\textbf{Step 7: Actionable Tactical Blueprint.}
Final recommendations spanned CTR/CR monitoring, creative iteration loops, CPM-sensitive bidding strategies, funnel-stage retargeting, and time-based budget optimization—each reflecting principles of advanced media planning.

These outputs demonstrate a key shift: LLMs are not merely automating reporting—they are simulating the interpretive, diagnostic, and strategic reasoning of human marketing analysts. Importantly, this reasoning was grounded in multimodal data and framed in the language of marketing practice, not engineering. The limitations are notable—some creative suggestions were generic, highlighting the need for tighter grounding in brand context. Yet for junior marketers or lean teams, such co-pilots represent a leap in decision support, converting raw metrics into actionable strategy at unprecedented speed.

In effect, GenAI enables a transition from retrospective analytics to real-time campaign sensemaking. This is not just automation — it is the first step toward cognitive collaboration between AI and a performance marketer.

\section{Conclusion and Future Work}

In this work, we introduced \somon, an explainable and generative AI agent framework designed to co-author marketing strategy with human teams. By fusing CTR-based content performance modeling with LLM-driven synthesis and reasoning, \somon bridges the persistent divide between unstructured advertising data and strategic actionability.

Crucially, \somon is not limited to insight generation. Through iterative content refinement guided by model feedback—and real-time campaign optimization suggestions informed by emergent performance signals—the system actively participates in the creative and tactical decision cycle. From predicting which creative assets will perform, to proposing rewrites, budget reallocations, or retargeting strategies, \somon demonstrates agent-like behavior in dynamically evolving campaign contexts.

Our deployments across agencies and global brands show that \somon does more than automate dashboards—it enables marketers to transition from manual execution to narrative intelligence. Tasks like persona mapping, content planning, and creative ideation—once requiring hours of human labor—are now accelerated up to 12$\times$, while retaining transparency through embedded explainability mechanisms.

\textbf{Future Work.} Looking ahead, two core directions shape the evolution of \somon:

\begin{enumerate}
    \item \textbf{Grounding Narrative Insights in Real Audience Data.} We plan to integrate statistically robust audience intelligence from databases such as Nielsen)~\cite{nielsen2025} and Global Web Index (GWI)~\cite{gwi2025}, aligning LLM-inferred personas and content narratives with empirically observed psychographics and behavioral signals. This will ensure that strategic recommendations are not only semantically coherent but also \textit{representative of actual audiences on the ground}.

    \item \textbf{Attributing Paid Media Performance to Content Pillars.} Another frontier involves correlating performance metrics across the paid media funnel—such as CPM, CTR, CR, and ROAS—with distinct content pillars (e.g., emotional appeal, product value, urgency). This will enable marketers to trace which narrative elements are truly driving business outcomes, establishing a framework for explainable attribution that connects story structure with media efficiency.
\end{enumerate}

More than a tool, \somon represents a new paradigm of co-creative AI — where strategic reasoning, storytelling, and optimization converge in a single loop. As a decision partner, not just an analyst, it marks a step toward the future of marketing—where human creativity is not replaced, but augmented, iterated, and grounded through intelligent collaboration with AI~\cite{karnatak2025acai}.

\section{AI Futures in Advertising: Replacement, Empowerment, or Insight?}

Recent industry discussions, led by Meta CEO Mark Zuckerberg, envision a radical future where advertisers provide only goals and budgets, while AI systems autonomously manage the entire advertising pipeline - creative, targeting, optimization, and measurement. Zuckerberg described this as a “redefinition of the category of advertising,” promising speed, scalability, and simplicity. However, critics warn of risks to transparency and trust, as such systems could displace agencies and in-house teams, leaving brands with little oversight of the “black box” outcomes \cite{forbes2025}.

Alternative approaches highlight different philosophies of AI in marketing. Google’s ACAI project, developed with DeepMind and Oxford researchers, positions AI as a co-creator: advertisers upload assets, insights, and inspiration, which are synthesized into structured “super-prompts” guiding campaign generation. This model enhances user agency and transparency, positioning AI as a creative partner rather than a replacement \cite{acai2025}. Similarly, Meta’s AI Sandbox explores collaborative generative tools that assist with copy, visuals, and resizing, while preserving human input. Empirical evidence suggests such co-created campaigns achieve significantly higher returns on ad spend, underscoring the value of AI-human collaboration \cite{fastcompany2025}.

A third path, represented by platforms such as SOMIN, Nielsen, and GWI, prioritizes insight and explainability over automation. Instead of generating ads, these systems analyze high-performing campaigns to surface patterns in audience targeting, creative direction, and content optimization \cite{_www25, nielsen2025, gwi2025}. This paradigm augments, rather than supplants, human creativity, embedding AI as an interpretive layer that grounds strategic decisions in data.

Together, these models illuminate three diverging philosophies: (i) \textit{replacement}, where AI supplants creative and strategic functions (Meta); (ii) \textit{empowerment}, where AI scaffolds human creativity (Google); and (iii) \textit{insight}, where AI explains and enhances strategic thinking (SOMIN \cite{Farseev2025SOMIN} and peers).

For agencies, the implications are profound \cite{Farseev2025Agencies}. Routine tasks such as asset resizing and A/B testing are increasingly automated, but the enduring value of agencies lies in orchestrating these AI systems, ensuring brand alignment, cultural nuance, and authenticity. Far from being rendered obsolete, agencies must evolve into facilitators and curators of AI-driven creativity, blending human storytelling with machine efficiency. In this sense, AI is not eliminating the role of agencies, but transforming them into higher-order strategists in the advertising ecosystem.

The accelerating adoption of AI across advertising is driving a profound shift in the skills required of agency professionals. While routine tasks such as asset resizing, reporting, and basic copywriting are increasingly automated, demand is growing for hybrid skillsets that combine human creativity with technical fluency. Industry leaders emphasize that traditional storytelling remains indispensable, but must now be paired with AI literacy, data interpretation, and prompt engineering \cite{creativeSalon2025,4As2025}. Many agencies are investing heavily in upskilling programs - for example, WPP reported over 150{,}000 AI training sessions for staff in 2025 - to ensure creative, media, and strategy teams can collaborate effectively with generative systems \cite{creativeSalon2025}. As a result, roles such as ``creative technologists,'' AI trainers, and ethics officers are emerging, reshaping agency talent profiles.

These shifts have fueled the growth of hybrid agencies that integrate creative storytelling, data science, and engineering capabilities within a single structure. Hybrid firms such as Accenture Song, which invested billions in AI talent and platforms, demonstrate how combining consulting-scale technical expertise with creative execution can disrupt traditional agency models. Analysts describe these players as ``talent magnets,'' attracting professionals eager to work across disciplines and leverage proprietary AI-driven solutions \cite{hybridAgencies2023,accentureSong2025}. For clients, this model offers end-to-end capabilities - from developing brand-trained language models to orchestrating cross-channel campaigns - that were previously fragmented across multiple agencies. Consequently, agencies that fail to evolve risk losing both clients and top talent to more agile, hybrid competitors.

Overlaying these transformations is an intensifying global competition for AI-savvy creative talent. Meta, in particular, has escalated the talent war by offering unprecedented compensation packages, sometimes exceeding \$100 million for senior AI researchers, along with access to cutting-edge computing infrastructure \cite{wired2025,reuters2025}.

Meta’s push to automate creative production threatens traditional agency roles, forcing agencies to stand out through culture, talent growth, and creative freedom. Since they cannot match big tech’s resources, their future lies in becoming AI facilitators and curators of authenticity, keeping human creativity central in an increasingly automated, platform-driven ecosystem.

\subsection*{Acknowledgements}
This work was supported by a grant provided by the Ministry of Economic Development of the Russian Federation in accordance with the subsidy agreement (agreement identifier 000000C313925P4G0002) and the agreement with the Ivannikov Institute for System Programming of the Russian Academy of Sciences dated June 20, 2025, No. 139-15-2025-011.


\bibliographystyle{ACM-Reference-Format}
\balance
\bibliography{main}

\end{document}